\documentclass[a4paper,11pt]{article}

\usepackage{amssymb,latexsym}

\begin{document}
\topmargin -10mm
\oddsidemargin 0mm

\renewcommand{\thefootnote}{\fnsymbol{footnote}}
\newcommand{\nn}{\nonumber\\}
\begin{titlepage}
\begin{flushright}
hep-th/0702029
\end{flushright}

\vspace*{10mm}
\begin{center}
{\Large \bf Thermodynamic Interpretation of Field Equations at
 Horizon of BTZ Black Hole}

\vspace*{20mm}

{\large M. Akbar~\footnote{Email address: akbar@itp.ac.cn}}\\
\vspace{8mm} { \em Institute of Theoretical Physics, Chinese
Academy of Sciences,\\
 P.O. Box 2735, Beijing 100080, China}

\end{center}

 \vspace{20mm}
 \centerline{{\bf{Abstract}}}
 \vspace{5mm}
A spacetime horizon comprising with a black hole singularity acts
like a boundary of a thermal system associated with the notions of
temperature and entropy. In case of static metric of BTZ black
hole, the field equations near horizon boundary can be expressed
as a thermal identity $dE = TdS + P_{r}dA$, where $E = M$ is the
mass of BTZ black hole, $dA$ is the change in the area of the
black hole horizon when the horizon is displaced infinitesimally
small, $P_{r}$ is the radial pressure provided by the source of
Einstein equations, $S= 4\pi a$ is the entropy  and $T = \kappa /
2\pi$ is the Hawking temperature associated with the horizon. This
approach is studied further to generalize it for non-static BTZ
black hole and show that it is also possible to interpret the
field equation near horizon as a thermodynamic identity $dE = TdS
+ P_{r}dA + \Omega_{+} dJ$, where $\Omega_{+}$ is the angular
velocity and $J$ is the angular momentum of BTZ black hole. These
results indicate that the field equations for BTZ black hole
possess intrinsic thermodynamic properties near horizon.

\end{titlepage}

\newpage
\renewcommand{\thefootnote}{\arabic{footnote}}
\setcounter{footnote}{0} \setcounter{page}{2}

%=================================sec.1 =================
%\sect{Introduction}

The Einstein's theory of general relativity predicts the existence
of black hole singularity comprising with spacetime horizons.
These horizons act as the boundaries of the spacetimes defined
mathematically, locally undetectable and block any physical
information to flow out to the rest of world. This led Bekenstein
\cite{a1} to claim that the black holes must hold non-zero entropy
since they withhold information from the outside observer. This
interpretation became unambiguous when Hawking \cite{a2} showed
that a black hole can emit thermal radiation with a temperature
proportional to its surface gravity at the black hole horizon and
with an entropy proportional to its horizon area \cite{a3} .The
Hawking temperature $T = \kappa / 2\pi$ and the black hole entropy
$S = A / 4G$ are connected through the identity $TdS = dE$,
usually called first law of black hole thermodynamics
\cite{a1,a2,a3}. In more general, the first law of black hole
thermodynamics is related with the energy change by
\begin{equation}
dE = TdS + work terms.
\end{equation}
In equation (1), the "$workterms$" have been provided differently
depending upon the type of black hole. For Kerr-Newman  black hole
family, the first law of black hole thermodynamics is given by
\begin{equation}
dE = TdS + \Omega dJ + \Phi dQ,
\end{equation}
where $\Omega = \frac{\partial M }{\partial J}$ is the angular
velocity and $\Phi = \frac{\partial M }{\partial Q}$ is the
electric potential. Equation (1) indicates that the thermodynamic
interpretation of the Einstein equations is possible near horizon
\cite{a5} because black hole solutions are derived from the
Einstein's field equations and the geometric quantities of the
spacetime metric are related with thermodynamic quantities. In
1995, Jacobson \cite{jac} was indeed able to find Einstein
equations by employing the first law of thermodynamics $\delta Q =
TdS$ along with proportionality of entropy to the horizon area of
the black hole. Recently, in case of Einstein gravity \cite{a6},
as well as for a more general Lanczos-Lovelock theories of
gravity, Paranjape Sarkar and Padmanabhan~\cite{PSP} found that it
is possible to interpret the field equations for a special class
spherically symmetric  as a thermodynamic identity $TdS = dE +
PdV$ near black hole horizon. For a more general situation in
Einstein gravity, Kothawala, Sarkar, and Padmanabhan \cite{ksp}
found that the field equations near any spherical symmetric
horizon can be expressed as a thermal identity $TdS = dE + PdV$
and extended their approach for stationary axis-symmetry horizons
and time dependent evolving horizon and found, in both cases, near
horizon structure of the Einstein equations can also be expressed
as a thermodynamic identity under the virtual displacement of the
horizon. Moreover, Hayward \cite{a8} derived $dE = TdS + WdV$,
called unified first law of black hole dynamics and relativistic
thermodynamics in spherically symmetric space-times, where $W$ is
the work density defined by $W = -\frac{1}{2}T^{ab}h_{ab}$.
However, it was soon realized that the notions of temperature and
entropy can be associated with several other types of horizon
which advocate a generic thermodynamic behavior of horizons
\cite{a11,a12,a13,a14}. In case of apparent horizon of FRW
universe, Cai and Kim \cite{a9} are able to derive the Friedmann
equations of (n+1)-dimensional Friedman-Robertson-Walker (FRW)
universe with any spatial curvature by applying the first law of
thermodynamics ($TdS=-dE$) to the apparent horizon. Also by using
the entropy expression of a static spherically symmetric black
hole in the Gauss-Bonnet gravity and in more general Lovelock
gravity, they reproduce the corresponding Friedmann equations in
each gravity. The possible extensions to the scalar-tensor gravity
and $f(R)$ gravity theory have been studied in reference
\cite{a10}. However, in more general context, the author of
present work and Cai \cite{a15} have shown  that thermodynamic
interpretation of Friedmann equations describing the dynamics of
the universe is possible at apparent horizon of FRW universe and
showed that it is possible to interpret the differential form of
Friedmann equations near apparent horizon as a universal form $dE
= TdS + WdV$ in the Einstein's theory of general relativity as
well as for a wider class of Gauss-Bonnet and Lovelock theories of
gravity, where $E$ is the total matter energy, $V$ is the volume
inside the apparent horizon, $P$ is the pressure of the perfect
fluid in the universe and $W$ is the work density. A possible
thermodynamic interpretation near horizons of field equations
within f(R) gravity for spherically symmetric black hole spacetime
as well as for a cosmological spacetime is studied in reference
\cite{a16}. These thermodynamic interpretation of gravitational
dynamics at horizons needs further investigation for understanding
it at a deeper level \cite{a5,a6}. In recent years,
(2+1)-dimensional BTZ (Banados-Teitelboim-Zanelli) black holes
have drawn a lot of attention as simplified models for exploring
conceptual issues relating black hole thermodynamics (see,
e.g.,\cite{btz, btz1}). The BTZ black hole is a solution of the
(2+1)-dimensional Einstein theory with a negative cosmological
constant $\Lambda = -1 / \ell^{2}$. The BTZ black hole solution
\cite{a17} is given by
\begin{equation}
ds^{2} = -f(r)dt^{2} + \frac{dr^{2}}{f(r)} + r^{2}(d\phi -
\frac{J}{2 r^{2}}dt)^{2},
\end{equation}
where the lapse function
\begin{equation}
f(r) = (-M + \frac{r^{2}}{\ell^{2}} + \frac{J^{2}}{4 r^{2}})
\end{equation}
involves two constants of integration $M$ and $J$ which represent
the mass and angular momentum of the BTZ black hole respectively.
Thus, for any value $\Lambda$ of the cosmological constant, there
is a 2-parameter family of BTZ metrics, characterized by $M$ and
$J$. The metric coefficient $f(r)$ vanishes at $r = r_{\pm}$,
where the outer and inner horizons are
\begin{equation}
r_{\pm} = \ell \left(\sqrt{M \pm \sqrt{M -
\frac{J^{2}}{\ell^{2}}}}\right) / \sqrt{2}.
\end{equation}
The 2-surface $r = r_{+}$ is specified as the black hole horizon
and exists only for $M > 0$ and $|J| < M \ell$. It is a null
stationary 2-surface. The Killing vector normal to this surface is
$n^{\alpha} = t^{\alpha} + \Omega_{+}\phi^{\alpha}$, where the
quantity $\Omega_{+}$ is interpreted as the angular velocity on
the horizon. It is convenient to express the mass $M$ and the
angular momentum $J$ in terms of $r_{+}$ and $r_{-}$
\begin{equation}
M = \frac{r_{+}^{2}+ r_{-}^{2}}{\ell^{2}},~~~~~~~~ J =
\frac{2r_{+}r_{-}^{2}}{\ell}.
\end{equation}
Here the units are such that $G_{3} = \frac{1}{8}$. The
Bekenstein-Hawking entropy of the BTZ black holes is twice the
perimeter of the event horizon \cite{a17}
\begin{equation}
S = 4\pi r_{+},
\end{equation}
the Hawking temperature and angular velocity can be computed
\begin{equation}
T = \frac{1}{4\pi}\frac{df(r)}{dr}|_{r=r_{+}},~~~~ \Omega_{+} =
\frac{J}{2r_{+}^{2}}
\end{equation}
These thermodynamic quantities obey the first law of
thermodynamics $dM = TdS + \Omega_{+}dJ$. Another characteristic
of the BTZ black hole is that its heat capacity $C_{J} =
(\frac{\partial M}{\partial T})_{J}$ is positive definite
\cite{hct}. This implies the temperature increases with the mass.
Hence, the BTZ black hole can be stable in thermal equilibrium
with any arbitrary volume of heat bath.

The purpose of this letter is to present a possible thermodynamic
interpretation of field equations for BTZ black hole near horizon.
In order to achieve this,I replace $f(r) = (-M +
\frac{r^{2}}{\ell^{2}} + \frac{J^{2}}{4 r^{2}})$ by an arbitrary
function $f(r)$ and extract a possible thermodynamic
characteristic of field equations near the horizon. Thus the
general static BTZ metric is given by
\begin{equation}
ds^{2} = -f(r)dt^{2} + \frac{dr^{2}}{f(r)} + r^{2} d\phi^{2}
\end{equation}
Let's assume that the function $f(r)$ has a simple zero at $r = a$
and $f^{\prime}(a)\neq 0$ but has a finite value at $r = a$. This
defines a space-time horizon at $r = a$ associated with a non-zero
surface gravity $\kappa = \frac{1}{2}f'(a)$ and temperature $T =
\kappa / 2\pi$ . The associated entropy of the horizon is
determined by the location of the zero at $r = a$ of the function
$f(r)$ and is proportional to the horizon area. The components of
the Einstein tensor $G_{ab} = R_{ab} - \frac{1}{2}R g_{ab}$ for
the metric (9) are given by
\begin{equation}\label{4}
G^{0}_{0} = G^{1}_{1} = \frac{f'(r)}{2r},
\end{equation}
and
\begin{equation}\label{6}
G^{2}_{2} = \frac{f''(r)}{2},
\end{equation}
where prime stands for the derivative with respect to $r$, the
subscripts a, b run from 0 to 2, $R_{ab}$ is the Ricci tensor in
(2+1)-dimensional Einstein's general relativity and $R$ is a Ricci
scalar. It can be seen readily that the components $G^{0}_{0}$ and
$G^{1}_{1}$ of Einstein tensor are equal and their evaluation at
$r = a$, give
\begin{equation}\label{7}
G^{0}_{0}|_{r=a} = G^{1}_{1}|_{r=a} = \frac{f'(a)}{2a}.
\end{equation}
From (2+1)-dimensional Einstein field equations
\begin{equation}
G_{ab} + \Lambda g_{ab} = - \pi T_{ab},
\end{equation}
it can be seen readily from equation (12) that the metric (9) will
satisfy (2+1)-dimensional Einstein field equation (13) provided
the stress energy tensor $T_{ab}$ has the form $T^{0}_{0}=
T^{1}_{1} = P_{r}$, where $P_{r}$ is the radial pressure provided
by the source. The Einstein equation for this metric when
evaluated at $r = a$ reads
\begin{equation}
\frac{1}{2a} f'(a) - \frac{1}{\ell^{2}} = -\pi T^{0}_{0}.
\end{equation}
This equation(14) describes the dynamics of gravity near horizon
of static BTZ black hole. My aim is to show that the above
equation possesses intrinsic thermodynamic properties and can be
expressed as a thermodynamic identity. To achieve this, let's
consider a virtual displacement $da$ of the horizon and then
multiply by it on the both sides of this equation(14). One can
rewrite it in the form
\begin{equation}
\frac{f'(a)}{4\pi} d(4\pi a) - \frac{d(a^{2})}{\ell^{2}} =
-P_{r}d(\pi a^{2}),
\end{equation}
where $T^{0}_{0} = P_{r}$ is a radial pressure of the source. One
can recognize the term $\frac{f'(a)}{4\pi}$ on the left side of
this equation as a temperature T and the quantity inside
parentheses on the left side is an entropy $S$ of BTZ black hole
associated with the horizon. The second term on left side of the
above equation is the change of mass $dM$. If one identifies the
mass $M = \frac{a^{2}}{\ell^{2}}$ of the BTZ black hole as a
energy $E$, the above equation can be written as
\begin{equation}
dE = TdS + P_{r}dA,
\end{equation}
which is the exact form of the first law of thermodynamic for BTZ
black hole. Where $dA$ is the change in horizon area and the term
$P_{r}dA$ corresponds to work done against the pressure. Hence,
the field equations near horizon of static metric for BTZ black
hole can be expressed as a thermodynamic identity under the
virtual displacement of the horizon.

Let's now turn to the case of non-static BTZ black hole metric(3).
This metric has two singularities $r_{\pm}$ comprising with outer
and inner event horizons. One can write the thermodynamic
quantities associated with horizon in terms of horizon radii
$r_{\pm}$. In order to study a possible thermodynamic
interpretation of field equations near horizon, one has to replace
$f(r) = (-M + \frac{r^{2}}{\ell^{2}} + \frac{J^{2}}{4 r^{2}})$ by
a generic $f(r)$ determined via (2+1)-dimensional Einstein field
equation(13). Thus, non-static BTZ metric with a generic $f(r)$
can be written as
\begin{equation}
ds^{2} = -f(r)dt^{2} + \frac{dr^{2}}{f(r)} + r^{2} (d\phi -
\frac{J}{2 r^{2}}dt)^{2}.
\end{equation}
Let's assume that the function $f(r)$ has two zeros at $r =
r_{\pm}$ and the outer horizon is an event horizon of BTZ black
hole associated with the notions of temperature of entropy. The
temperature $T$, entropy $S$, and angular momentum $J$ associated
with the horizon located at $r = r_{+}$ are  $T =
\frac{1}{4\pi}f'(r)|_{r = r_{+}}$, $S = 4\pi r_{+}$  and $J
=2\Omega  r_{+}^{2}$ respectively. Using (2+1)-dimensional
Einstein field equation (13), one can write (1, 1)-component of
the field equation in the form
\begin{equation}\label{5}
\frac{J^{2} + 2r^{3}f'(r)}{4r^{4}} - \frac{1}{\ell^{2}} = -\pi
T^{1}_{1},
\end{equation}
where $T^{1}_{1}$ is the (1,1)-component of the stress-energy
tensor which corresponds to the radial pressure $P_{r}$ of the
source at horizon. The above equation when evaluated at horizon $r
= r_{+}$, can be written as
\begin{equation}\label{6}
\frac{J^{2}}{4 r_{+}^{3}} + \frac{1}{2}f'(r_{+}) -
\frac{r_{+}}{\ell^{2}} = - P_{r}(\pi r_{+}).
\end{equation}
The above equation(19) describes the dynamic of the spacetime near
horizon. My aim is to determine a possible thermodynamic
interpretation of this equation(19) near horizon. For this
purpose, consider a virtual displacement $dr_{+}$ of the horizon
and then multiply by it on both sides of equation(19). The
resulting equation can be rewritten as
\begin{equation}\label{7}
\frac{J^{2}}{2 r_{+}^{3}}dr_{+} + \frac{f'(r_{+})}{4\pi} d(4\pi
r_{+}) - \frac{2 r_{+}dr_{+}}{\ell^{2}} = - P_{r}(2\pi
r_{+}dr_{+})
\end{equation}
This equation(20) can be simplified further by substituting $dM =
\frac{2r_{+} dr_{+}}{\ell^{2}} + 2 \Omega_{+}^{2} r_{+}dr_{+}$ in
it to get
\begin{equation}
4 \Omega_{+}^{2} r_{+}dr_{+} + \frac{f'(r_{+})}{4\pi} d(4\pi
r_{+}) - dM = - P_{+}d(\pi r_{+}^{2}).
\end{equation}
We know that $J = 2 \Omega r_{+}^{2}$ and by taking its
differential it is trivial to write
 \begin{equation}
 dJ = 4\Omega_{+} r_{+}dr_{+}.
 \end{equation}
By substituting the value of $dJ$ in equation (21), it is easy to
see that the first term on the left of equation(21) turn out to be
$\Omega_{+} dJ$, while the second term on the left side of this
equation(23) is trivial to recognize as $TdS$. If one identifies
mass $M$ of the BTZ black hole as a energy $E$, the above equation
can be written as
\begin{equation}
 dE = TdS + P_{r}dA + \Omega dJ,
\end{equation}
which is identical to the first law of thermodynamics for
no-static BTZ black hole. Hence the field equations near horizon
of non-static BTZ black hole behave like a thermal system
satisfying the first law of thermodynamics.

In summary, It has shown that thermodynamic interpretation of
field equations for BTZ black hole is possible near horizon. In
case of static metric of BTZ black hole, the field equation near
horizon can be rewritten as a thermodynamic identity, $dE = TdS +
P_{r}dA$, where $E = M$ is the mass inside the horizon of BTZ
black hole, $P_{r}$ is the radial pressure of the source and $A$
is the area enclosed by the horizon. It has also shown that the
field equations for non-static BTZ black hole can be expressed as
a first law $dE = TdS + P_{r}dA + \Omega dJ$ of black hole
thermodynamics. It is interesting to further investigate the
present work to the charged BTZ black hole case. These results are
certainly associated with the holographic properties of gravity.
It would be of great interest to analyze further the consequences
of these observations to the holographic principle.
\section*{Acknowledgments}
I would like to thanks Rong-Gen Cai for his useful discussion and
comments. The work was supported in part by a grant from Chinese
Academy of Sciences, by NSFC under grants No. 10325525 and No.
90403029.

%\end{twocolumn}

\end{document}